\documentclass[preprint,nofootinbib,APS]{revtex4}
\usepackage{amsmath,amssymb}
\usepackage{bm}

\begin{document}

\title{
Self-Duality Equations on $S^6$ from $\mathbb{R}^7$ monopole
}
\author{Hironobu Kihara and 
Eoin \'O Colg\'ain}

\preprint{KIAS-P09036}

\affiliation{Korea Institute for Advanced Study\\
207-43 Cheongnyangni 2-dong, Dongdaemun-gu, 
Seoul 130-722, Republic of Korea
}

\date{\today}

\begin{abstract}
In this note we identify a correspondence between a seven-dimensional monopole configuration 
of the Yang-Mills-Higgs system and the generalized
self-dual configuration of the Yang-Mills system on a six-dimensional sphere. 
In particular, the topological charge of the self-duality configurations belongs to 
the sixth homotopy group of the coset $G/H$ associated with the symmetry breaking $G \rightarrow H$ induced by a non-trivial Higgs configuration in seven-dimensions. 
\end{abstract}

\maketitle


In this short note we make an observation about the self-duality equations on the six-dimensional sphere. We make use of the work of 
\cite{Tchrakian:1978sf, Bais:1985ns, Saclioglu:1986qn, Kihara:2007di}, the details of which we omit. 
It is well known \cite{Nakahara:1990th} that a four-dimensional instanton configuration has second Chern character, which is in turn, related to the third homotopy group $\pi_3(G)$ of the gauge group $G$. We show there is a correspondence between seven-dimensional monopoles and self-duality equations on the six-dimensional sphere. { There have been numerous efforts to generalize monopoles to higher dimensions, some of which have appeared in \cite{Tchrakian:1978sf, Yang:1977qv, Ito:1984wu, 
Chen:1998qb, Kihara:2008rv, Diaz:2008kd}. } 

In analogy in six-dimensions, when $G=SU(N)$, the third Chern character ${\rm Tr} F^3$ is considered as a topological charge and takes values in $\pi_5(G)$, with $\pi_5(SU(N)) = \mathbb{Z}$ for $N \geq 3$. In particular, for $SU(4) \simeq SO(6)$\footnote{This is easily embedded in $SU(N)$ with $N \geq 4$.} pure Yang-Mills theory on $S^6$, 
one has a non-trivial gauge configuration \cite{Kihara:2007di}, which satisfies the generalized self-duality relation
\begin{align}
\label{eqn:self-dual}
c F \wedge F &= *_6 F. 
\end{align}
Here, $c = 3/(\bm{q} R_0^2)$ is a covariantly constant scalar given in terms of the gauge coupling $\bm{q}$ and radius of $S^6$ $R_0$. 

A few examples of
other configurations for $\pi_5(G) \neq 0$ have appeared in the literature in \cite{Saclioglu:1986qn}. In this note, our focus is non-trivial solutions of self-duality equations on $S^6$ with gauge group $G$ with $\pi_5(G)= 0$. 

In one dimension higher, the above equation takes the form
\begin{align}
F \wedge F &= *_{7}  \tilde{c} \{ D \phi , F \},
\label{eqn:seven-bogomolnyi}
\end{align}
where $\tilde{c}$ is a constant. The above equation can { be} obtained from the Bogomol'nyi equation \cite{Kihara:2008rv}. 
Here $F$ is a gauge field strength two-form and ``$*_7$" is the Hodge dual operator with respect to the Euclidean metric on $\mathbb{R}^7$. 
$\phi^a$ are scalar fields forming a fundamental multiplet of SO(7), 
$\phi := \phi^a \gamma_a$ and finally, $D$ is the covariant exterior derivative: $D \phi = d \phi + g [A , \phi]$. 
The Hermitian matrices $\gamma_a$, $(a=1,2,\cdots, 7)$, are Dirac matrices with respect to SO(7), with 
$\gamma_{ab}:= (1/2)[\gamma_a , \gamma_b ]$ satisfying the commutation relations of SO(7). $\phi$ induces symmetry breaking when it acquires an expectation value $ \| \langle \phi^a \rangle \| = H_0$. 

To substantiate this connection, we suppose that the gauge configuration is concentrated around the origin of $\mathbb{R}^7$ .
Solutions of Eq.~(\ref{eqn:seven-bogomolnyi}) 
represent monopole configurations with corresponding topological charge, 
\begin{align}
Q &= \int_{B(R_0)} {\rm Tr} D \phi F^3 = \int_{S^6_{R_0}} {\rm Tr} \phi F^3 ~,
\end{align}
where $B(R_0)=\{ x \in \bm{R}^7 | \| x \| \leq R_0 \}$. 
This charge $Q$ relates to the mapping class degree of $S^6_{R_0} \rightarrow SO(7)/SO(6) =S^6$ for the case where $R_0 >> 1$. 
To see this, we suppose that gauge field $A$ and scalar field $\phi$ have the following form, 
\begin{align}
A &= \frac{1-K(r)}{2{\bm{q}}} ede ~,&
\phi &= H_0 U(r) e ~,&
e &= \frac{x^a}{r} \gamma_a~,
\end{align}
where $\bm{q}$ is again the gauge coupling, $r = \sqrt{x_{a} x^{a}}$ and the functions $U(r)$ and $K(r)$ satisfy the following boundary conditions: $U(0)=1, K(0)= 1, U(\infty)=1$ and $K(\infty)=0$.
The corresponding $F$ and $D \phi$ are
\begin{align}
F &= \frac{1-K^2}{4{\bm{q}}} de \wedge de - \frac{K'}{2{\bm{q}}} e dr \wedge de~,&
D \phi &= H_0 ( KU de +U' edr )~.
\end{align}
For this particular configuration, Eq.~(\ref{eqn:seven-bogomolnyi}) reduces to a first order nonlinear ordinary differential equation \cite{Kihara:2008rv}. 

In the asymptotic region, $F$ and $D \phi$ become
\begin{align}
F & \rightarrow \frac{1}{4{\bm{q}}} de \wedge de~,& 
D \phi & \rightarrow  H_0  U' edr ~,
\label{eqn:}
\end{align}
where, as may be seen, $F$ is aligned perpendicular to the radial direction and thus, along the $S^6$. 
Hence $F$ can be regarded as a differential form on $S^6$. In this asymptotic region, 
Eq.~(\ref{eqn:seven-bogomolnyi}) is tranformed into Eq.~(\ref{eqn:self-dual}) with a suitable scalar. 

However, the above discussion includes some degree of approximation: the self-duality is not exact. If we now relax the constraint of demanding a finite energy configuration by considering the singular configuration 
\begin{align}
A &= \frac{1}{2{\bm{q}}} ede~, &\phi &=  - \frac{\kappa}{r}  e~,
\end{align} 
where $\kappa$ is a constant, the seven-dimensional equation 
\begin{align}
F \wedge F &= *  {\bf i} \mu   \{  D\phi, F  \}~, &
\mu &= \frac{3}{2 \bm{q} \kappa}~,
\end{align}
reduces to Eq.~(\ref{eqn:self-dual}). 


Having constructed a concrete example, we now consider other embeddings. In general, we may consider a gauge group $G$ with non-trivial $\pi_6(G/H)$ with 
symmetry breaking $G \rightarrow H$, from 
a seven-dimensional monopole solution. 
For simplicity suppose that $G$ is a simple group and the rank of group $G$ is greater than or equal to 3. 
From the long exact sequence of homotopy group 
we obtain 
\begin{align}
\pi_6 (G/H) \simeq {\rm Ker} \{ \pi_5(H) \rightarrow \pi_5(G)  \}~. 
\end{align}
If $\pi_5(G)=0$ and $H$ includes Spin(6) or SU($N$) $(N \geq 3)$ as a factor group, then $\pi_6(G/H) \neq 0$.

In contrast to the earlier example where the Higgs is in the fundamental $\mathbf{7}$ of $SO(7)$, it is possible to embed it and the adjoint $\mathbf{21}$ of $SO(7)$ in the adjoint $\mathbf{28}$ of $SO(8)$. Here $\pi_6(G/H) \neq 0$, and we can embed the above solution into the larger gauge theory with adjoint Higgs field and it does not come loose { as a result of a gauge transformation of the larger group}. 
E$_8$, SU($N$), $(N \geq 8)$ and SO$(N)$ $(N \geq 8)$ also permit
the same configuration with adjoint Higgs. 
 It would be interesting to explore embeddings of this configuration in string theory or M-theory: the gauge groups $SO(16)$ and $E_8$, both appear in \cite{Horava:1995qa}.
For example, it may be possible to consider symmetry breakings $SO(16) \rightarrow SO(6) \times SU(5) \times U(1)$ and $E_8 \rightarrow SU(4) \times SU(5) \times U(1)$, 
inspired by the symmetry breaking of  $SO(10)$ GUT: $SO(10) \rightarrow SU(5) \times U(1)$.  

For these symmetry breakings $\pi_6(G/H)\neq 0$.  
It may also be of interest to consider coupling this system to gravity in a similar fashion to studies appearing in \cite{Gibbons:2006wd, Brihaye:2006xc, Kihara:2009ea}, the latter of which addresses the possibility of cosmological models as a result of dynamical compactification on $S^6$.

 {\bf Acknowledgment}
HK is grateful to Muneto Nitta for his advice.  
We are also appreciate Jong-Chul Park and Jae-Yong Lee for their constructive comments. 
 

\end{document}